\documentclass[12pt]{article}
\usepackage{epsfig}
\usepackage{graphicx}
\usepackage{amssymb}

\def\beq{\begin{equation}}
\def\eeq#1{\label{#1}\end{equation}}
\def\eeqn{\end{equation}}

\def\beqa{\begin{eqnarray}}
\def\eeqa#1{\label{#1}\end{eqnarray}}
\def\eeqan{\end{eqnarray}}

\let\bar=\overbar

\def\Dslash{\not{\hbox{\kern-4pt $D$}}}
\def\dslash{\not{\hbox{\kern-2pt $\del$}}}

\def\msb{{\bar{\ssstyle M \kern -1pt S}}}

\usepackage{fancyhdr,graphicx}
\def\Title#1{\begin{center} {\Large {\bf #1} } \end{center}}

\begin{document}

\Title{Inhomogeneous seeding of quark bubbles in Neutron Stars}

\bigskip\bigskip


\begin{raggedright}

\center{\it   M. \'Angeles P\'erez-Garc\'ia}\\
\bigskip
Department of Fundamental Physics, University of Salamanca, \\
Plaza de la Merced s/n 37008 Salamanca, Spain\\

\end{raggedright}

\section{Introduction}

The question of the composition of the ultimate and densest phase of matter realized in nature is a fundamental problem in physics. Approaches to a partial answer have arisen from different directions. From the theoretical point of view, in the early universe the condensation from a quark phase to a nucleon phase has already been studied using cosmological arguments \cite{cosmo}. On the other side, at RHIC experimental efforts from the heavy-ion community trying to produce the so-called  quark-gluon plasma have yield even more questions on the nature of this phase of matter \cite{brook}. The possibility of quark deconfinement seems possible in a very dense and/or high temperature environment according to the quantum chromo dynamics (QCD) temperature-baryonic density phase boundary $T(\rho_B)$. In this context, it is especially interesting that the conditions believed to exist inside dense cold stars are  those matching the low-temperature, high-density region of the before-mentioned phase space. Neutron stars (NS) may constitute, in this way, possible sites where nucleon and quark phases can coexist and serve as a test-bench at the extreme for fundamental interactions. 

Besides standard-model baryonic matter, the presence of another type of matter present in the universe, dark matter (DM), may add interesting and new insights to current standard scenarios. This type of matter makes more than 80$\%$ of the current matter content of our universe and has been confirmed by precision measurements from a range of areas, for example, rotation curves in our galaxy, such as those recently presented in \cite{iocco}. The intimate nature of the dark matter remains to be determined, i.e. its fermionic or bosonic nature or its couplings with the rest of the standard-model particles. There is a huge world-wide effort to constrain its properties either in a direct, indirect or collider search. These missing pieces of information will be key in revealing the possible observable signatures of its existence.

\begin{figure}[htb]
\begin{center}\includegraphics[width=0.75\textwidth]{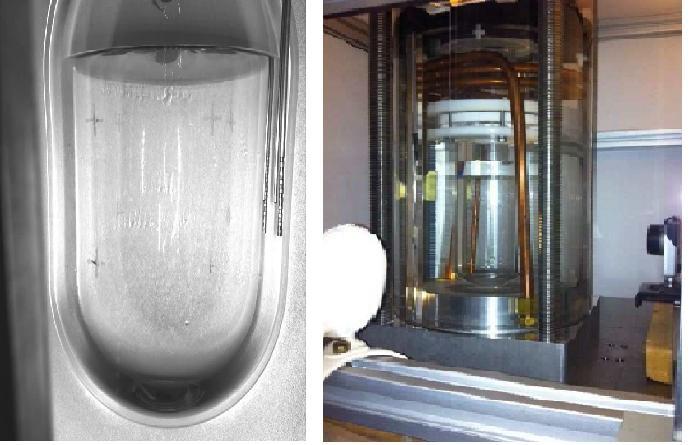}
\caption{  2L PICO bubble chamber (left panel) and  MOSCAB experiment based on the geyser technique (right panel). Figures courtesy of  both collaborations.}
\end{center}
\label{fig1}
\end{figure}

In this contribution we elaborate on  the consequences of the presence of DM in a dense nuclear matter star environment focusing on indirect effects, namely that of bubble nucleation. Let us mention that this same output is already considered as a {\it direct search} product in experimental settings with detected events consisting of recoiling of a nuclear target after a weakly interacting particle hits on it. Experiments along these lines include  PICO \cite{pico}, SIMPLE \cite{Simple} or MOSCAB \cite{moscab}, see Fig.(\ref{fig1}).

The consideration of dense star settings requires the scaling of the same type of physics used in bubble chambers or superheated droplet detectors in extreme environments with differences of several orders of magnitude in density and temperature. This is a  real theoretical challenge and requires state-of-the-art microscopic calculations at the quantum level. Although quantum bubble nucleation has already been discussed theoretically in the literature \cite{madsen} in the context of quantum fluctuations of temperature or seeding \cite{prl}, it remains to be tested on earth from the experimental point of view. The key to this process is the energy deposition that could lead confined quarks out of the nucleon bag, overcoming a $\sim (0.1-1)$ GeV potential energy barrier. As we will argue the presence of external energy sparks due to DM particle self-annihilations could indeed drive the deconfinement transition.
\begin{figure}[htb]
\begin{center} 
\includegraphics [scale =0.35]{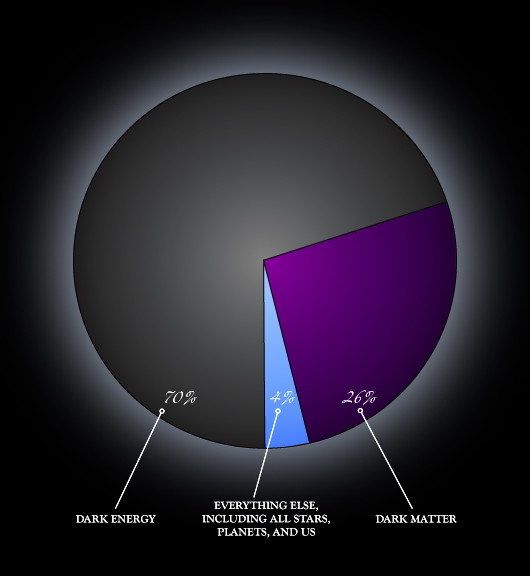}
\caption{Content of the current universe. Dark matter amounts to $23\%$. Figure from Chandra observatory website at http://chandra.harvard.edu}
\label{Fig1a}
\end{center}
\end{figure}

\section{Dark matter as a nucleation trigger source}

In the classical bubble nucleation theory of Seitz \cite{seitz} an initial localized energy deposition over a threshold value, $E_0$, in a dense medium, is the mechanism to drive microscopic bubble formation. Further modellization allows that a typical ionization reaction involving complex processes of atomic collision-cascades is replaced  by an abrupt temperature rise in an infinitesimal cylindrical volume around the ion trajectory at the time-of-passage. Activated particles can populate a metastable state with energy $E$ with a probability $\sim e^{-E/k_BT}$.  If the high temperature endures only for a short time, the higher states have only a small chance to return to the ground state, due to the potential-energy threshold associated.

In a NS core, the temperature and density conditions are those of a degenerate Fermi system. Typical nucleon energies are those of the Fermi level, fulfilling  the relationship $E/k_B T \simeq E_F/k_B T\gg1$. For such a system mass density is thought to be several times that of nuclear saturation, $\rho_0\simeq 2.8\times 10^{14}\,\rm g/cm^3$, and it is highly improbable to excite fluctuations of $\delta T\sim 10$ MeV estimated to be necessary for the quark deconfinement \cite{horvath}.

In this pictured scenario an external agent is needed to perturbe the nucleon degeneracy in order to allow the system to free some of the quark content. At this point we must include the dark matter fraction. It amounts to $23\%$ of the current content in our universe, 8 times larger than the standard-model particle  content, see a cartoon in Fig. (\ref{Fig1a}). Let us  consider a Majorana DM component, electrically neutral, capable of self-annihilating and massive enough to be gravitationally accreted from an existing galactic DM distribution onto a NS. At a given galactocentric location close to the sun the local DM density $\rho_{\chi}\rm (r)\sim 0.3\, GeV/cm^3$ is taken as reference. Particle candidates of this kind are indeed in the literature \cite{bertone}, one of the most popular, the neutralino, belongs to the supersymmetric class. For a  mass range $m_{\chi}\sim (10-10^4$) $\rm GeV/c^2$ and inside the NS, we will take a (spin independent) scattering cross section with nucleons around the sensitivity values of current direct detection experiments,  $\sigma_{\chi N}\sim 10^{-44}\,\rm cm^2$. This choice allows building a stabilized thermalized fraction of DM aside the standard baryonic component over time. Using this fact one can write the DM particle number density \cite{nusi} as 
\begin{equation}
n_\chi(r, T)=n_{0, \chi} e^{-\left(\frac{r}{r_{\rm th}}\right)^2},
\end{equation}
 where $r_{\rm th}$ is the thermal radius  $r_{\rm th}= \sqrt{\frac{3 k_B T}{2 \pi G \rho_n m_{\chi}}}$ and $n_{0, \chi}$ is the central DM density value normalized to the global DM population inside the star of mass $M$ and radius $R$ at a given time, $N_{\chi}(t)$.

We take DM self-annihilation as the only process removing DM from the star. The $\chi$-population number is obtained from the solution of an ordinary differential equation $\dot N_{\chi}=C_\chi-C_a N^2_{\chi}$. It explicitly includes competing processes from capture and annihilation with rates ${C_{\chi}}$ and  ${C_a}$ respectively \cite{gould1987}. The capture rate is given by 
\small
\begin{equation}
C_{\chi}\simeq \frac {3 \times 10^{22}}{f\left(\frac{M}{R}\right)} \left(\frac{M}{1.5 M_{\odot}}\right) \left(\frac{R}{10\,\rm km}\right) \left(\frac{1\, \rm TeV}{m_{\chi}}\right)\left(\frac{\rho_{\chi}}{0.3 \rm \frac{GeV}{cm^3}}\right)\,\,\rm s^{-1},
\label{capture}
\end{equation}
\normalsize
with $f\left(\frac{M}{R}\right)={1-0.4 \left(\frac{M}{1.5 M_{\odot}}\right) \left(\frac{10 \,\rm km}{R} \right)}$ a redshift correction factor. Following \cite{kouv} we do not consider a reduction of the cross-section since we use values well above the geometrical cross-section. It is important to note that this dependence sets a limit on the intrinsic capability of NS to accumulate a critical amount of DM and possibly serve as a test-bench for DM properties.
For the annihilation rate we write
\begin{equation}
C_a=\int n^2_\chi(r, T) \langle \sigma_a v\rangle\,dV .
\end{equation}
\noindent Then we can finally write
\begin{equation}
N_{\chi}(t)=\sqrt{\frac {C_{\chi}}{C_a}} \rm coth \left[\frac{(t-t_{\rm col})}{\tau}+coth^{-1} 
\left( \sqrt{\frac{C_a}{C_{\chi}}} N_{\chi}(t_{\rm col})\right) \right],
\label{eqnchi}
\end{equation}
with $\tau^{-1}=\sqrt{{C_{\chi}}{C_a}}$  the inverse relaxation time to achieve equilibrium and $N_{\chi}(t_{\rm col})$ is the number of  DM particles inside the progenitor core at the time of the collapse (NS birth). This population is essentially inherited from the progenitor star in its lifetime.

\section{Induced bubble nucleation}

The possibility to nucleate a quantum bubble of quark matter will be driven by the ability of the DM self-annihilation process to effectively inject the spark of energy needed to activate quarks confined in the nucleon bag. Note that the ultimate goal of this process could be the formation of a bubble, in a similar way to thermal boiling. Since there are however some slight differences among both processes, we will however assume here that the energy required is the same for the sake of brevity.

The so-called free energy of nucleation is approximately  \cite{madsen} given by $F\simeq \frac{4 \pi}{3}R^3(P_n-P_q)+8\pi \gamma R$ where the two terms account for the pressure and curvature energy contributions, respectively. In order to produce spherical bubbles the minimum of this quantity shows a critical radius, namely $\frac{\partial F}{\partial r}=0$ implies
\begin{equation}
R_c=\sqrt{\frac{2 \gamma}{\Delta P}},
\end{equation}
where $\Delta P=P_q-P_n$ and $P_q$ ($P_n$) is the quark (nucleon) pressure and stability is granted as the contained baryonic number $A\sim \frac{4}{3} \pi R^3_c n_n > A_{\rm min}$ and $A_{\rm min}\gtrsim 10$ \cite{Wilk96}. We will consider here, for simplicity, that the system is pure neutron matter.
 
In the approximation of very light quarks (i.e u,d quarks),  $m_q\simeq 0$ and creation of neutrally charged bubbles this critical work is given by
\begin{equation}
F(R_c)=W_c=\frac{16 \pi}{3} \sqrt{\frac{2 \gamma^{3}}{\Delta P}}.
\end{equation}
For a two-flavor ud-quark system in the MIT bag model  (and $B > (145\,\rm MeV)^4$) \cite{madsen}) is given by $P_{q}=\sum\limits_{i=u,d} \frac{\mu^4_i}{4\pi^2}-B$ and assuming a pure n system $P_n\simeq \frac{(\mu^2_n-m^2_n)^{5/2}}{15\pi^2m_n}$ as most of the pressure will effectively be provided by neutrons in the n fluid \cite{madsen}.  $\gamma=\sum\limits_{i=u,d} \frac{\mu^2_i}{8\pi^2}$, is the curvature coefficient and $\mu_{i}$ $(\mu_n$) is the quark (nucleon) chemical potential related to the Fermi momentum of the degenerate system $\mu_i=p_{F\,i}$ ($\mu_n=\sqrt{m^2_n+p^2_{F\,n}}$). Electrical charge neutrality requires for the ud matter number densities $n_d=2 n_u$ and $n_n=(n_u+n_d)/{3}$ with $n_i= \frac{\mu^3_i}{\pi^2}$ in the light quark massless limit.

\subsection{Activation of quarks}
In this subsection we grossly estimate the injection of energy from final state photonic channels coming from the DM self-annihilation. We will discard the many-body blocking in nuclear phase space, estimated to be a minor effect at this range of energies. Additional channels should be considered carefully, although large masses $m_\chi\gtrsim 1$ $\rm GeV/c^2$ allow a safe semiclassical treatment of the DM component. The rate of external energy (photon channels) from the annihilations can be written as
\begin{equation}
\frac{dE}{dt}=\int E Q(E, r, T) dE dV,
\end{equation}
where $Q(E,r)$ is a function defined as a sum for each possible annihilation channel as
\begin{equation}
Q(E, r, T)=\sum_i Q_{i,\chi} (E, r, T) \simeq n^2_\chi(r, T) \langle \sigma_a v\rangle \frac{dN^i_{\gamma}}{dE}.
\end{equation}
$\frac{dN^i_{\gamma}}{dE}$ is the photon spectrum for the i-th annihilation channel. Let us note that the low-energy region of the spectrum will not be capable of activating the quarks if $E<E_0$, then, an efficiency of the process, $f_a<1$, is expected. However, in terms of mass fractions for example, even a $f_a=0.1\%$ for a $m_\chi\sim$ TeV particle would allow the triggering.  It is worth mentioning at this point that, in principle, also decay modes could serve as an activation method \cite{decay}, however we will restrict ourselves to annihilation in this work.

\begin{figure}[htb]
\includegraphics[width=0.5\textwidth]{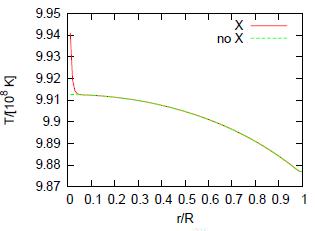}
\includegraphics[width=0.5\textwidth]{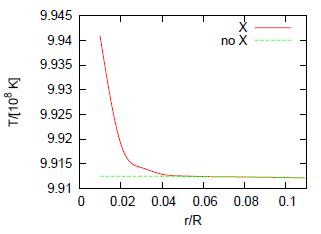}
\caption{ (Left panel) Internal temperature radial profile in a NS with a self- annihilating DM component with $m_\chi=10$ $\rm TeV/c^2$ at a $t=0.1$ yr since collapse.  (Right panel) Inner core range $\lesssim 1$ km \cite{lauria}. See text for details.}
\label{fig2}
\end{figure}

Fig.~(\ref{fig2}) panel on the left shows internal temperature vs. radius in the star. The local heating is compared in a standard (MURCA neutrinos \cite{page} and surface photon emission) scenario (solid line) and that  due to DM annihilation for a $m_\chi=10$ $\rm TeV/c^2$ (dashed line). On the right panel the local temperature radial profile is zoomed to the inner $\sim$1 km. Inner early heating could produce effective boiling. Details are given in \cite{lauria}. An arbitrary flat profile with $T=10^9$ K is supposed at initial time and a value $\langle \sigma_a v\rangle\simeq 3\times 10^{-26}\,\rm cm^3/s$ for the velocity averaged DM annihilation cross section is taken.

The time averaged energy density coming from this DM process can be estimated as
\begin{equation}
\langle u_{DM} \rangle_{\Delta t} \simeq \frac{1}{\Delta t}\int_{t_0}^{t_0+\Delta t} [ \int E Q(E, r) dE] dt.
\end{equation}
This is to be compared to the energies quoted to produce bubbles \cite{harko}, $u_{\rm bub} \simeq W_c/V_d\simeq 5.4\times 10^{35} \rm \,erg/cm^{3}$, or in more familiar nuclear units $u_{\rm bub}\sim 0.1 $ $\rm GeV/fm^3$. The spatial spread is $\Delta r \simeq \frac{\hbar c}{\Delta E}=\frac{\hbar c}{f_a m_\chi c^2} \simeq 2 \, 10^{(-4\div-1)} \,\rm fm$, ensuring the local character of the energy release. In heavy-ion collisions the estimated energy density for deconfinement is $u\gtrsim 5$ $\rm GeV/fm^3$ at low pressures \cite{qcd} or five times smaller in lattice QCD calculations \cite{lattice}. It seems that a high pressure environment could catalyze the reaction, although further refinement of the models is still needed.


\section{Bubble associated instabilities}

In this scenario the number of nucleated {\it stable} bubbles created can be accumulated  in the central region over a period of time. To produce an observable effect we expect that after a critical number of bubbles is created the coalescence of bubbles in the dense nucleon system may induce further macroscopic structure changes in the star. The quoted number for this to happen is however not clear at this point. Some authors like Harko and collaborators \cite{harko} state that one single bubble $N_0=1$ will be responsible for driving a NS to quark star (QS) transition. This is due to a bubble dynamical mechanism allowing it to grow to become comparable to the star size. More in general, the relationship to analyze is given by
\begin{equation}
{N}_{\rm bub} \simeq \int \frac{dN_{\rm bub}}{dE} \frac{dE}{dt} dt \geq N_0.
\nonumber
\end{equation}
In this contribution, we will be conservative and will study the consequences of driving a moderate mechanical instability in the central core of the star.
If this was the case the equation of state of baryonic matter $P(T, \rho)$ would allow pressure oscillations that could disrupt the star producing a conversion into a more compact object. Whether the conversion is partial or complete is not yet completely understood.

As a consequence of the triggering, the creation of a number of bubbles $\delta N_{\rm bub}$ could cause a perturbation of the hydrostatic pressure value 
\begin{equation}
\delta P\simeq \left[\frac{\partial P}{\partial N_{\rm bub}}\right]_0 \delta N_{\rm bub},
\end{equation}
with respect to the standard value. The derivative coefficient is related to regular nuclear matter compressibility (in the absence of bubbles).
To create an instability, this change should be such that the star can not adjust adiabatically to it so that a catastrophic event is driven. In order to quantify the effect, let us consider an old and cold NS with $T\sim 10^5$ K and $\rho_n= 5 \rho_0$ for a DM particle with $m_{\chi}\sim1$ TeV.
In principle, a standard variation in the number of nucleons depleted from the thermal volume nucleon sea, $\sqrt{A_{{\rm th}}}\simeq 10^{21}$, could provide a large enough variation in pressure to trigger transitioning. Then the number of bubbles to create this perturbation is $N_0\simeq \sqrt{A_{{\rm th}}}/A_{\rm min}\simeq 10^{20}$. 
From eq.(\ref{capture}) we can see than there is a comparable capture rate of DM particles into the star. At late times $t>>\tau$ equilibrium can be achieved and an efficient mechanism could indeed create such a number of bubbles, precipitating the macroscopic conversion of the nuclear star into a (hybrid) QS.
The conversion itself has been extensively studied in the literature see for example \cite{harko, bombaci, daigne} where emission of gamma ray burst is likely to follow the rapid (ms timescale) event. Since agile methods of gamma-ray detection are increasingly available, characteristic features in the photon emission predicted in these events (sharp signal) could help disentangle the formation of these compact objects and go deeper into the possible formation mechanisms in the future.


\subsection*{Acknowledgement}

I express my thanks to the organizers of the CSQCD IV conference for the invitation to participate and for providing an 
excellent atmosphere which was the basis for inspiring discussions with all participants. This research has been supported at University of Salamanca by the Consolider MULTIDARK and FIS2012-30926 MINECO projects.

\end{document}